\input{aipcheck}
\documentclass[,final,numberedheadings]{aipproc}
\usepackage{subfigure}
\usepackage{epsfig}
\layoutstyle{6x9}

\begin{document}
\title{Rapidity dependence of Bose-Einstein correlations at SPS energies}

\author{S.~Kniege       }       {address={\ORTi}}
\author{C.~Alt}                 {address={\ORTi}}
\author{T.~Anticic}             {address={\ORTu}}
\author{B.~Baatar}              {address={\ORTh}}
\author{D.~Barna}               {address={\ORTd}}
\author{J.~Bartke}              {address={\ORTf}}
\author{L.~Betev}               {address={\ORTj}}
\author{H.~Bialkowska}          {address={\ORTs}} 
\author{C.~Blume}               {address={\ORTi}}  
\author{B.~Boimska}             {address={\ORTs}}
\author{M.~Botje}               {address={\ORTa}}
\author{J.~Bracinik}            {address={\ORTc}} 
\author{R.~Bramm}               {address={\ORTi}} 
\author{P.~Bun\v{c}i\'{c}}      {address={\ORTi},altaddress={\ORTj}} 
\author{V.~Cerny}               {address={\ORTc}} 
\author{P.~Christakoglou}       {address={\ORTb}} 
\author{O.~Chvala}              {address={\ORTo}}
\author{J.G.~Cramer}            {address={\ORTq}} 
\author{P.~Csat\'{o}}           {address={\ORTd}} 
\author{P.~Dinkelaker   }       {address={\ORTi}}
\author{V.~Eckardt      }       {address={\ORTn}} 
\author{D.~Flierl       }       {address={\ORTi}} 
\author{Z.~Fodor        }       {address={\ORTd}} 
\author{P.~Foka }               {address={\ORTg}}                                                
\author{V.~Friese       }       {address={\ORTg}} 
\author{J.~G\'{a}l      }       {address={\ORTd}}
\author{M.~Ga\'zdzicki  }       {address={\ORTi},altaddress={\ORTl}}
\author{V.~Genchev}             {address={\ORTv}} 
\author{G.~Georgopoulos}        {address={\ORTb}} 
\author{E.~G{\l}adysz   }       {address={\ORTf}} 
\author{K.~Grebieszkow  }       {address={\ORTt}}
\author{S.~Hegyi        }       {address={\ORTd}} 
\author{C.~H\"{o}hne    }       {address={\ORTm}} 
\author{K.~Kadija       }       {address={\ORTu}} 
\author{A.~Karev        }       {address={\ORTn}} 
\author{M.~Kliemant     }       {address={\ORTi}} 
\author{V.I.~Kolesnikov }       {address={\ORTh}} 
\author{E.~Kornas       }       {address={\ORTf}} 
\author{R.~Korus        }       {address={\ORTl}} 
\author{M.~Kowalski     }       {address={\ORTf}} 
\author{I.~Kraus        }       {address={\ORTg}} 
\author{M.~Kreps        }       {address={\ORTc}} 
\author{A.~Laszlo       }       {address={\ORTd}} 
\author{M.~van~Leeuwen  }       {address={\ORTa}} 
\author{P.~L\'{e}vai    }       {address={\ORTd}} 
\author{L.~Litov        }       {address={\ORTr}} 
\author{B.~Lungwitz     }       {address={\ORTi}}
\author{M.~Makariev     }       {address={\ORTr}} 
\author{A.I.~Malakhov   }       {address={\ORTh}} 
\author{M.~Mateev       }       {address={\ORTr}} 
\author{G.L.~Melkumov   }       {address={\ORTh}}
\author{A.~Mischke      }       {address={\ORTg}} 
\author{M.~Mitrovski    }       {address={\ORTi}} 
\author{J.~Moln\'{a}r   }       {address={\ORTd}} 
\author{St.~Mr\'owczy\'nski}    {address={\ORTl}}
\author{V.~Nicolic}             {address={\ORTq}}
\author{G.~P\'{a}lla    }       {address={\ORTd}} 
\author{A.D.~Panagiotou}        {address={\ORTb}} 
\author{D.~Panayotov    }       {address={\ORTr}}
\author{A.~Petridis     }       {address={\ORTb}} 
\author{M.~Pikna        }       {address={\ORTc}} 
\author{D.~Prindle     }       {address={\ORTk}}
\author{F.~P\"{u}hlhofer}       {address={\ORTm}}
\author{R.~Renfordt     }       {address={\ORTi}} 
\author{C.~Roland       }       {address={\ORTe}} 
\author{G.~Roland       }       {address={\ORTe}}
\author{M. Rybczy\'nski}        {address={\ORTl}} 
\author{A.~Rybicki      }       {address={\ORTf},altaddress={\ORTj}}                              
\author{A.~Sandoval     }       {address={\ORTg}} 
\author{N.~Schmitz      }       {address={\ORTn}} 
\author{T.~Schuster     }       {address={\ORTi}}
\author{P.~Seyboth      }       {address={\ORTn}}
\author{F.~Sikl\'{e}r   }       {address={\ORTd}} %
\author{B.~Sitar        }       {address={\ORTc}} 
\author{E.~Skrzypczak   }       {address={\ORTt}}
\author{G.~Stefanek     }       {address={\ORTl}}
\author{R.~Stock        }       {address={\ORTi}} 
\author{C.~Strabel      }       {address={\ORTi}}
\author{H.~Str\"{o}bele }       {address={\ORTi}} 
\author{T.~Susa         }       {address={\ORTu}}
\author{I.~Szentp\'{e}tery}     {address={\ORTd}} 
\author{J.~Sziklai      }       {address={\ORTd}}
\author{P.~Szymanski    }       {address={\ORTl},altaddress={\ORTl}}
\author{V.~Trubnikov    }       {address={\ORTl}}
\author{D.~Varga        }       {address={\ORTd}} 
\author{M.~Vassiliou    }       {address={\ORTb}}
\author{G.I.~Veres      }       {address={\ORTd},altaddress={\ORTe}}
\author{G.~Vesztergombi }       {address={\ORTd}}
\author{D.~Vrani\'{c}   }       {address={\ORTg}} 
\author{A.~Wetzler      }       {address={\ORTi}}
\author{Z.~W{\l}odarczyk}       {address={\ORTl}}
\author{J.~Zim\'{a}nyi  }       {address={\ORTd}}
\newcommand {\ORTa} {NIKHEF, Amsterdam, Netherlands. }
\newcommand {\ORTb}{Department of Physics, University of Athens, Athens, Greece.}
\newcommand {\ORTc}{Comenius University, Bratislava, Slovakia.}
\newcommand {\ORTd}{KFKI Research Institute for Particle and Nuclear Physics,Budapest,Hungary.}
\newcommand {\ORTe}{MIT, Cambridge, USA.}
\newcommand {\ORTf}{Institute of Nuclear Physics, Cracow, Poland.}
\newcommand {\ORTg}{Gesellschaft f\"{u}r Schwerionenforschung (GSI), Darmstadt, Germany.}
\newcommand {\ORTh}{Joint Institute for Nuclear Research, Dubna, Russia.}
\newcommand {\ORTi}{Fachbereich Physik der Universit\"{a}t, Frankfurt, Germany.}
\newcommand {\ORTj}{CERN, Geneva, Switzerland.}
\newcommand {\ORTk}{University of Houston, Houston, TX, USA.}
\newcommand {\ORTl}{Institute of Physics \'Swi{\,e}tokrzyska Academy, Kielce, Poland.} 
\newcommand {\ORTm}{Fachbereich Physik der Universit\"{a}t, Marburg, Germany.}
\newcommand {\ORTn}{Max-Planck-Institut f\"{u}r Physik, Munich, Germany.}
\newcommand {\ORTo}{Institute of Particle and Nuclear Physics, Charles University, Prague,Czech Republic.}
\newcommand {\ORTp}{Department of Physics, Pusan National University, Pusan, Republic ofKorea.}
\newcommand {\ORTq}{Nuclear Physics Laboratory, University of Washington, Seattle, WA, USA.}
\newcommand {\ORTr}{Atomic Physics Department, Sofia University St. Kliment Ohridski, Sofia,Bulgaria.} 
\newcommand {\ORTs}{Institute for Nuclear Studies, Warsaw, Poland.}
\newcommand {\ORTt}{Institute for Experimental Physics, University of Warsaw, Warsaw,Poland.}
\newcommand {\ORTu}{Rudjer Boskovic Institute, Zagreb, Croatia.}
\newcommand {\ORTv}{Institute for Nuclear Research and Nuclear Energy,Sofia, Bulgaria.} 
\begin{abstract}
This article is devoted to results on $\pi^{-}$-$\pi^{-}$-Bose-Einstein 
correlations in central Pb+Pb collisions measured by the NA49 experiment at the CERN SPS. 
Rapidity as well as transverse momentum dependences of the correlation lengths will be   
shown for collisions at 20\itshape A\upshape, 30\itshape A\upshape, 40\itshape A\upshape, 80\itshape A\upshape, and 158\itshape A \upshape GeV beam energy.
Only a weak energy dependence of the radii is observed at SPS energies.
The $k_t$-dependence of the correlation lengths as well
as the single particle $m_t$-spectra will be compared to model calculations.
The rapidity dependence is analysed in a range of 2.5 units of rapidity starting at 
the center of mass rapidity at each beam energy. The correlation lengths measured in 
the longitudinally comoving system show only a weak dependence on rapidity.
\end{abstract}
\maketitle
\section[Introduction]{Introduction}
The measurement of correlations of identical bosons in heavy ion collisions provides a unique
tool to investigate the space time evolution of the particle emitting source.
Bose-Einstein correlations are observed as an enhancement of the
yield of pairs of particles with small relative momenta. Measurements of the range and
strength of the correlations in momentum space allow to derive the extension
of the source in coordinate space. Due to space momentum correlations in expanding sources
the correlations do not reflect the whole extensions of the source.  
In such a scenario, the study of the correlations in different regions of phase
space helps to understand the evolution of the source.
While the dependence of the correlation lengths on the mean transverse momentum 
$k_{t}=\frac{1}{2}|\vec{p}_{t,1}+\vec{p}_{t,2}|$ of the pairs
reflects the transverse expansion
dynamics of the source the dependence on the pair rapidity 
$Y=\frac{1}{2}\log\left(\frac{E_{1}+E_{2}+p_{z,1}+p_{z,2}}{E_{1}+E_{2}-p_{z,1}-p_{z,2}}\right)$, 
which is measured in the center of mass system, should shed light on the profile of the
source in longitudinal direction.
The large acceptance of the NA49 experiment allows us to obtain a comprehensive
picture of the dynamical evolution of the source.\\[0.2cm]
The article is organized as follows: A brief survey of the experiment, the
construction of the correlation function and the fit method are presented in
section \ref{sec:exp}. In section \ref{sec:sys}, crucial systematic uncertainties 
in the analysis due to detector effects are discussed. 
The results on the 
$k_{t}$- as well as the $Y$- dependence of the correlation lengths are
presented in section \ref{sec:res} and compared to model calculations.
\section{Experimental setup and analysis}
\label{sec:exp}
\begin{figure}[]
\centering
\begin{minipage}[t]{15.0cm}
\centering
\includegraphics[width=0.75\textwidth]{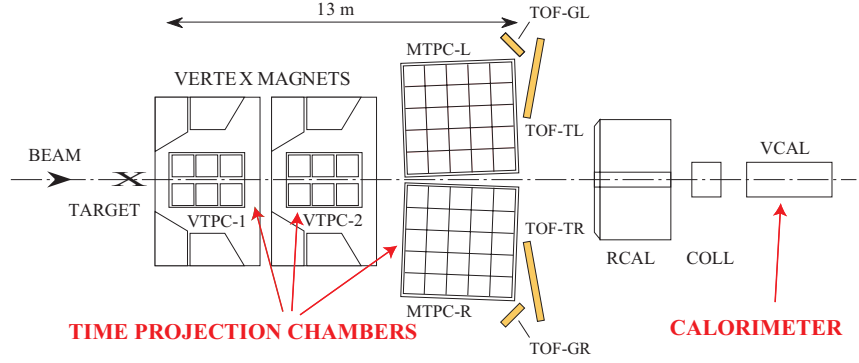}
\caption{NA49 detector setup.}
\label{fig:setup}
\end{minipage} 
\end{figure}

NA49 \cite{A.1} is a fixed target experiment located at the CERN SPS comprising
 four large-volume Time Projection Chambers (TPC), two of which are 
located inside the magnetic field of two superconducting dipole magnets (Figure \ref{fig:setup}). 
The TPCs are read out at 90 (MTPC) and 72 (VTPC) pad rows resulting in a very good
determination of the momentum of the traversing particles.
A zero degree calorimeter at the downstream end of the experiment is used to trigger on
the centrality of the collisions. 
The data presented here correspond to the 7.2\% most central events for data samples taken at 20\itshape A\upshape,
30\itshape A\upshape, 40\itshape A\upshape, 80\itshape A\upshape, and 158\itshape A \upshape GeV beam energy.
By scaling the magnetic field it was possible to obtain a similar coverage of
phase space relative to the center of mass rapidity for the different beam energies.
Measuring the specific energy loss dE/dx of charged particles in the
gas of the TPCs with a resolution of 3-4\% allows particle identification. 
However, due to ambiguities in particle identification by specific energy loss
measurements in certain regions of phase space, negative hadrons rather than identified
negative pions are studied in this analysis.\\
The correlation function is constructed as the ratio of a distribution of the momentum
difference of pairs from the same event (signal) and a mixed event background distribution (background).
Following the approach of 
Pratt and Bertsch \cite{A.2,A.4} the momentum difference is decomposed into a component
parallel to the beam axis
$q_{long}$ and two components in the transverse plane $q_{out}$ and $q_{side}$ with
$q_{out}$ defined parallel, and $q_{side}$ perpendicular to $k_{t}$.
The correlation function is parameterised by a Gaussian function
\begin{eqnarray}
C_{2}(q)_{BP}=1+\lambda\cdot\exp(-R_{out}^{2}q_{out}^{2}-R_{side}^{2}q_{side}^{2}-R_{long}^{2}q_{long}^{2}-2R_{outlong}^{2}q_{out}q_{long})
\label{gl:BP}
\end{eqnarray}
and the parameters $R_{out}$, $R_{side}$, $R_{long}$, $R_{outlong}$, and $\lambda$ are
determined by a fit to the measured correlation function.
The Coulomb repulsion of the particles is accounted for by weighting the theoretical
Bose-Einstein correlation function $C_{2}(q)_{BP}$  by a factor $F(q_{inv}$,<$r$>$)$ \cite{A.3}
in the fit procedure.
The weight is determined by the invariant momentum difference $q_{inv}$ of the pair and the mean
pair separation <$r$> of the particles in the source. According to \cite{A.3} this quantity can be derived from the
extracted radii. We therefore determine the source parameters as well as the value of
<$r$> in an iterative fit-procedure. 
For following fit function was used:
\begin{eqnarray}
C_{2}(q)_{f}= n\left\lbrace  p\cdot(C_{2}(q)_{BP}\cdot F(q,<r>))+(1-p) \right\rbrace .
\label{gl:fit}
\end{eqnarray} 
The contamination of the sample with pairs of non-identical particles and 
pairs of pions from long lived resonances or weak decays is accounted for by 
a purity factor $p$ which is determined by a VENUS/GEANT
simulation. The fit parameter $n$ is introduced to account for the different statistics
in signal and background distributions.
Beside the uncertainties which arise due to the construction of the correlation function
and the fit formalism there are further detector related effects which will be
discussed in the next section.
\section{Systematic studies}
\label{sec:sys}
\subsection{Two track resolution}
The momentum difference of a pair is closely related to the distance of the
tracks traversing the detector. 
Bose-Einstein correlations are restricted to a narrow
window in momentum difference, hence it is crucial to understand the two track
resolution of the detector. The overlap of charge clusters induced at the pad planes of
the TPCs can lead to an assignment of points to the wrong track or in an extreme case to
complete merging of two tracks. In this case, a pair with small relative momentum will be
lost in the signal distribution and the observable Bose-Einstein enhancement will be reduced. 
To study the impact of the limited two track resolution on the extracted radii, the distance of 
the tracks was measured at each pad row where both tracks lie in the sensitive volume of the TPCs.
\begin{figure}[b!]
\centering
\begin{minipage}[]{16.8cm}
\centering
\includegraphics[width=0.9\textwidth]{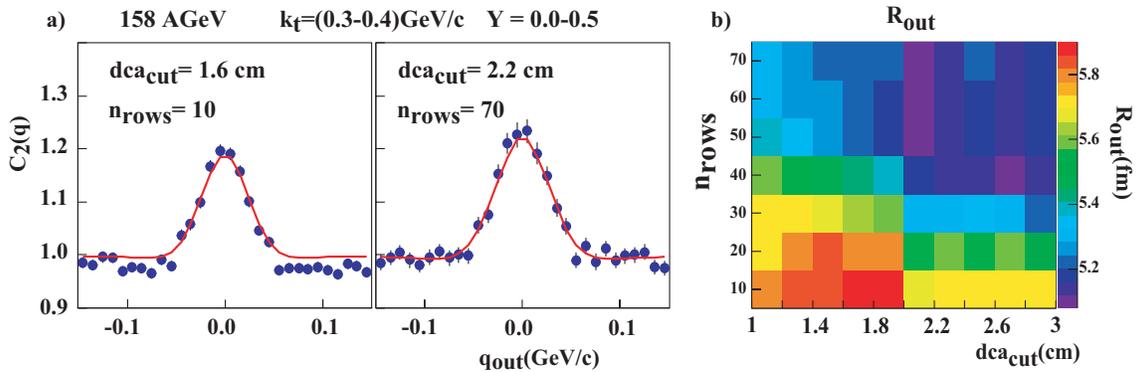}
\caption{a) Impact of two track resolution inefficiencies on the shape of the correlation function and b) fit results for $R_{out}$ 
for different combinations of the cut parameters $dca_{cut}$ and $n_{rows}$.}
\label{fig:dca_study}
\end{minipage} 
\end{figure}
Starting from the downstream end of the TPCs 
the distance of closest approach ($dca$) of two tracks after a given number of 
passed pad rows $n_{rows}$ was determined. Pairs with a $dca$ smaller than a given cut value $dca_{cut}$
were rejected both from the signal and the background distribution. The impact of a variation of 
the two parameters $dca_{cut}$ and $n_{rows}$ on the correlation function is shown in Figure \ref{fig:dca_study}.
Requiring only small values of $dca_{cut}$ and $n_{rows}$, tracks can approach each other very closely 
over a considerable part of the track length in the TPCs. In this case, track merging effects can lead to 
a significant loss of pairs in the signal.
This effect is very pronounced at high transverse momenta and shows up in an undershoot of the 
projection of the correlation function onto $q_{out}$ (Figure \ref{fig:dca_study}a). 
Increasing the cut parameters, the influence of merging 
effects is reduced, the undershoot of the correlation function vanishes 
and the extracted radii vary only by less than 0.2 fm (Figure \ref{fig:dca_study}b).
This can serve as an estimate of the systematic error on the radii due to the specific treatment of the
two track inefficiencies in this analysis.\\[0.6cm]
Further systematic errors on the radii arise due to uncertainties concerning the treatment of the Coulomb interaction (which can 
significantly influence the parameters $R_{out}$ and $\lambda$), the missing particle identification, 
the finite momentum resolution 
of the detector 
and the normalisation of the correlation function. The overall systematic error on the radii is not specified for each 
$k_{t}$-$Y$-bin but estimated to be smaller than 1 fm for all extracted radii. 
\section{Results}
\label{sec:res}
\subsection{The $k_t$-dependence}
\label{sec:kt}
\begin{figure}[]
\begin{centering}
\begin{minipage}[]{14.5cm}
\centering
\includegraphics[width=1.0\textwidth]{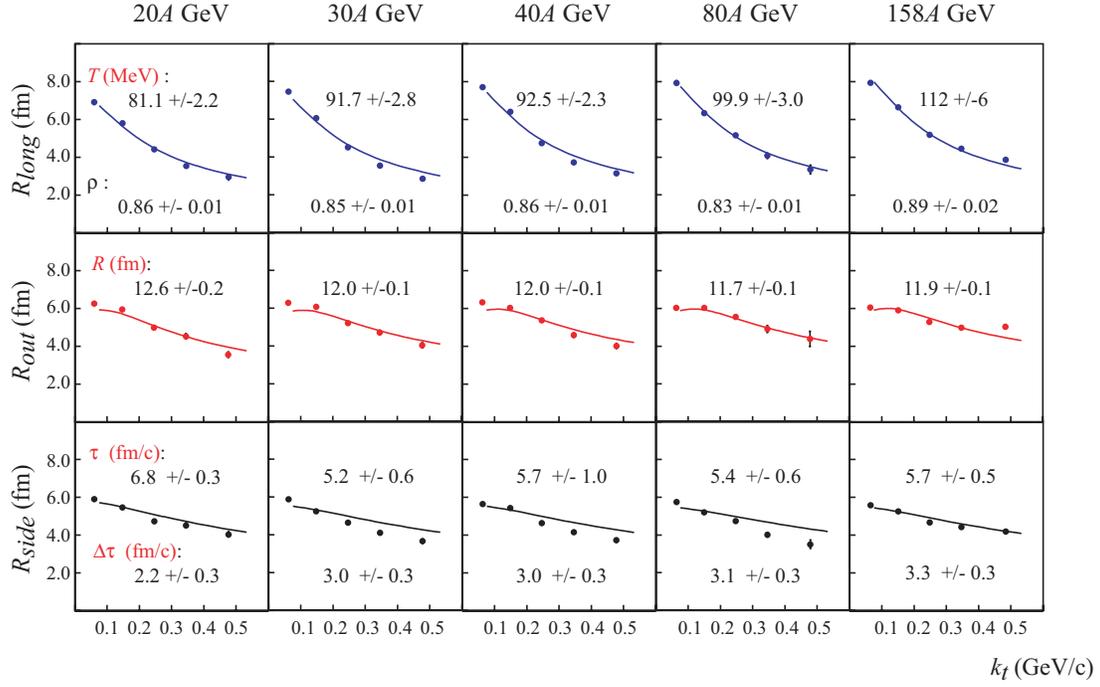}
\caption{The $k_{t}$-dependence of $R_{side}$, $R_{out}$, and $R_{long}$ at midrapidity ($0.0<Y<0.5$) for the different data sets (dots).
The lines correspond to a combined fit of the blast wave model to the radii and the single particle $p_{t}$-spectra. }
\label{fig:ktdep}
\end{minipage} 
\end{centering}
\end{figure}

Figure \ref{fig:ktdep} presents the $k_t$-dependence of the radii at midrapidity.
Radius and $\lambda$ parameters were obtained from fits of \eqref{gl:fit} to the correlation function in bins of $Y$ and $k_t$. 
Since the influence of the finite momentum resolution was found to be small, no corrections were applied for this effect. 
The bin width in $k_{t}$ was chosen to be 0.1 GeV/c and increased to 0.2 GeV/c for the 
last bin ((0.4-0.6) GeV/c) to obtain sufficient statistics at all energies.  
A strong decrease of $R_{long}$ with $k_{t}$ is observed. $R_{out}$ is slightly larger 
than $R_{side}$ for all energies indicating a finite duration of particle emission \cite{A.4}.
Also shown in this picture is a fit of the $k_{t}$-dependence of the radii according to a blast wave parameterisation \cite{A.5}
of the source. In this model the source is treated as boost invariant in the longitudinal direction. In the transverse
direction a box-shaped density profile and a linearly increasing flow profile is assumed. 
Space momentum correlations induced by flow reduce the measured correlation lengths. This effect is partly compensated 
in case of a superimposed thermal velocity field.  
Therefore ambiguities arise in the two model parameters temperature and flow, which can not be resolved by only analysing the $k_{t}$-dependence of the radii.
To resolve these ambiguities the single particle $p_{t}$-spectra of protons and negatively charged pions measured by NA49 were fitted
to a parameterisation derived from the same model.
The lines in Figure \ref{fig:ktdep} correspond to a combined fit to the radii and the particle $p_{t}$-spectra.
As expected from the weak energy dependence of the radii only small variations of the extracted source parameters were observed.
The extracted parameters were the temperature $T$, the maximum transverse flow rapidity $\rho$, the transverse geometrical radius $R$, the emission time $\tau$ and the emission duration $\Delta\tau$. 
The fit results are inserted in Figure \ref{fig:ktdep}. The temperature $T$ slightly increases with the beam energy, transverse flow and geometrical radius stay approximately constant over the observed energy range. For the emission time we obtain values of 5.4 to 6.8 fm/c. 
The fit slightly overpredicts $R_{side}$ at high $k_{t}$ but still results in a finite emission duration of 2.2-3.2 fm/c.
The fit might be further constrained by adding more particle spectra or by including contributions from resonance decays 
in the model.
\subsection{$Y$-dependence}
\label{sec:y}
\begin{figure}[]
\centering
\begin{minipage}[]{14.5cm}
\centering
\includegraphics[width=1.0\textwidth]{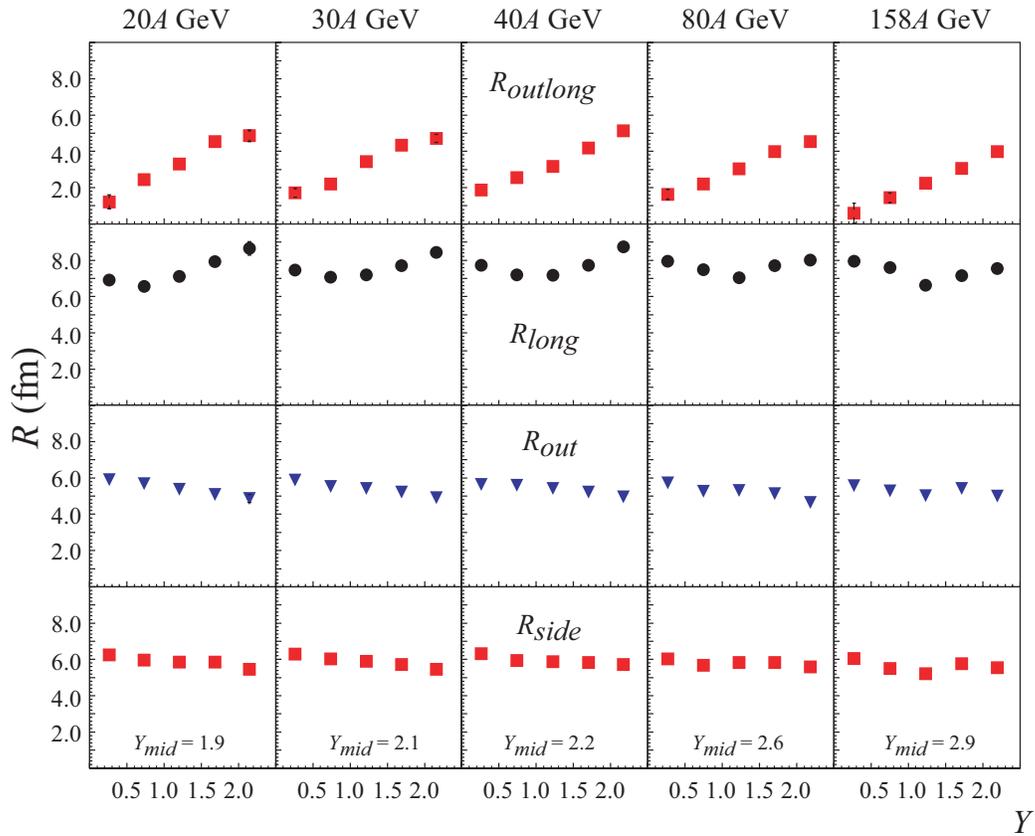}
\caption{Rapidity dependence of $R_{side}$, $R_{out}$, $R_{long}$, and $R_{outlong}$ at $k_{t}$=(0.0-0.1) GeV/c for the different beam energies. Shown are as well the midrapidity values $Y_{mid}$ for the different beam energies.}
\label{fig:rapdep}
\end{minipage} 
\end{figure}

The model described in section \ref{sec:kt} is only applicable in case of a longitudinally boost invariant source. 
Under such conditions it is expected that the cross term $R_{outlong}$ vanishes \cite{A.6}. 
Considering the systematic error of 1 fm this condition is fulfilled to good approximation at midrapidity.
In Figure \ref{fig:rapdep} the rapidity dependence of $R_{long}$, $R_{side}$, $R_{out}$, and $R_{outlong}$ is shown at $k_{t}$=(0.0-0.1) GeV/c
for the different beam energies.
In \cite{A.7} the impact of a non-boost invariant expansion on the parameter $R_{outlong}$ is studied.
An increase of $R_{outlong}$ with increasing rapidity is predicted 
due to the decrease of the inclusive pion yields and is in agreement with the results 
for all energies.
However a change in the longitudinal expansion dynamics which is indicated by the change in $R_{outlong}$ 
is not reflected in the rapidity dependence of the other observables. $R_{side}$, which is supposed to determine the geometrical size of the 
source \cite{A.8} does not change significantly with rapidity. $R_{out}$ is approximately constant over the investigated rapidity region. Only slight changes in 
$R_{long}$ are observed. These are even less pronounced at higher transverse momenta.\\[0.1cm]
In summary,
a distinct energy dependence of the radii is not observed
even though there is a dramatic change in the energy dependence 
of other hadronic observables like e.g. the kaon to pion ratio \cite{A.9}.
Furthermore 
the radii do not show a pronounced rapidity dependence, 
in contrast to the particle yields which decrease strongly with
rapidity.
\begin{theacknowledgments}
This work was supported by the US Department of Energy
Grant DE-FG03-97ER41020/A000,
the Bundesministerium fur Bildung und Forschung, Germany, 
the Virtual Institute VI-146 of Helmholtz Gemeinschaft, Germany,
the Polish State Committee for Scientific Research (1 P03B 097 29, 1 PO3B 121 29,  2 P03B 04123), 
the Hungarian Scientific Research Foundation (T032648, T032293, T043514),
the Hungarian National Science Foundation, OTKA, (F034707),
the Polish-German Foundation, the Korea Research Foundation Grant (KRF-2003-070-C00015) and the Bulgarian National Science Fund (Ph-09/05).
\end{theacknowledgments}

\end{document}